\documentclass[pdflatex,sn-mathphys-num]{sn-jnl}

\usepackage{graphicx}%
\usepackage{multirow}%
\usepackage{amsmath,amssymb,amsfonts}%
\usepackage{amsthm}%
\usepackage{mathrsfs}%
\usepackage[title]{appendix}%
\usepackage{textcomp}%
\usepackage{manyfoot}%
\usepackage{booktabs}%
\usepackage{algorithm}%
\usepackage{algorithmicx}%
\usepackage{algpseudocode}%
\usepackage{listings}%
\usepackage{adjustbox}
\usepackage{caption}
\usepackage{booktabs} 
\usepackage{tabularx} 
\usepackage{tikz} 
\definecolor{Red}{RGB}{203, 65, 84}
\definecolor{Blue}{RGB}{20, 20, 200}
\definecolor{Celadon}{RGB}{195, 208, 180}
\usepackage{array}

\theoremstyle{thmstyleone}%
\theoremstyle{thmstyletwo}%

\theoremstyle{thmstylethree}%

\begin{document}

\title[Article Title]{The Kansei Engineering Approach in Web Design:
Case of Transportation Website}

\author[1]{\fnm{Alisher} \sur{Akram}}

\author[1]{\fnm{Aray} \sur{Kozhamuratova}}

\author*[1]{\fnm{Pakizar} \sur{Shamoi}}\email{p.shamoi@kbtu.kz}

\affil*[1]{\orgdiv{School of Information Technology and Engineering}, \orgname{Kazakh-British Technical University}, \orgaddress{\street{Tole bi Street 59}, \city{Almaty}, \postcode{050000}, \country{Kazakhstan}}}








\abstract{
Kansei Engineering (KE) is a user-centered design approach that emphasizes the emotional aspects of user experience. This paper explores the integration of KE in the case of a transportation company that focuses on connecting cargo owners with transportation providers. The methodology involves aligning the design process with the company's strategy, collecting and semantic scaling Kansei words, and evaluating website design through experimental and statistical analyses. Initially, we collaborated with the company to understand their strategic goals, using Use Case and Entity Relationship diagrams to learn about the website functionality. Subsequent steps involved collecting Kansei words that resonate with the company's vision. Website samples from comparable transportation companies were then evaluated by X subject in the survey. Participants were asked to arrange samples based on emotional feedback using a 5-point SD scale. We used Principal Component Analysis (PCA)  to identify critical factors affecting users' perceptions of the design. Based on these results, we collaborated with designers to reformulate the website, ensuring the design features aligned with the Kansei principles. The outcome is a user-centric web design to enhance the site's user experience.  This study shows that KE can be effective in creating more user-friendly web interfaces in the transportation industry.



}




\maketitle

\section{Introduction}

In today's digital era, web interfaces facilitate user interactions and enhance user experiences. The design of web interfaces significantly impacts users' perceptions, satisfaction, and engagement with online platforms.

Kansei is a Japanese term that refers to a customer's psychological response to a product\cite{Simon2004}. Kansei can be referred to as sensitivity, sensibility,
feeling, and emotion \cite{Nagamachi1999}. So, it is an emotion somebody gets from a certain product using all her senses of vision, feeling, and taste. Kansei Engineering (KE) is an Affective Engineering method focusing on product development or improvement while considering customers' psychological feelings and expectations\cite{Nagamachi1999}. Kansei Design, in turn, refers to KE-based works that produce actual products for industry \cite{ecd}.

Researchers in cognitive psychology, user experience, interaction design, Kansei design, and design of emotions, industrial designers, engineers, and ergonomists can benefit from KE \cite{Mahut2017}. Integrating KE with human-computer cognitive psychology principles can lead to better website design and evaluation, resulting in improved interactive trading effects\cite{bian2018}.

\begin{figure}[h]
    \centering
    \includegraphics[width=0.8\textwidth]{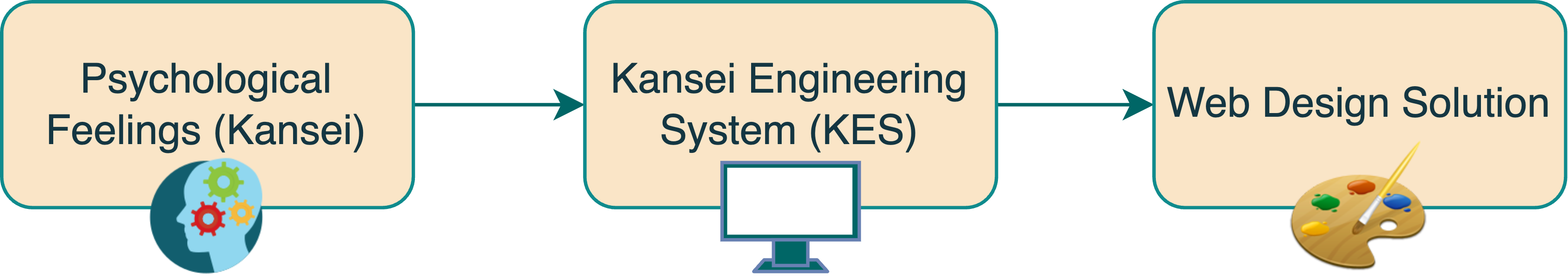}
    \caption{The general idea of  Kansei Engineering applied to Website Design}
    \label{kanseiidea}
\end{figure}

\begin{figure}[h]
    \centering
    \includegraphics[width=0.8\textwidth]{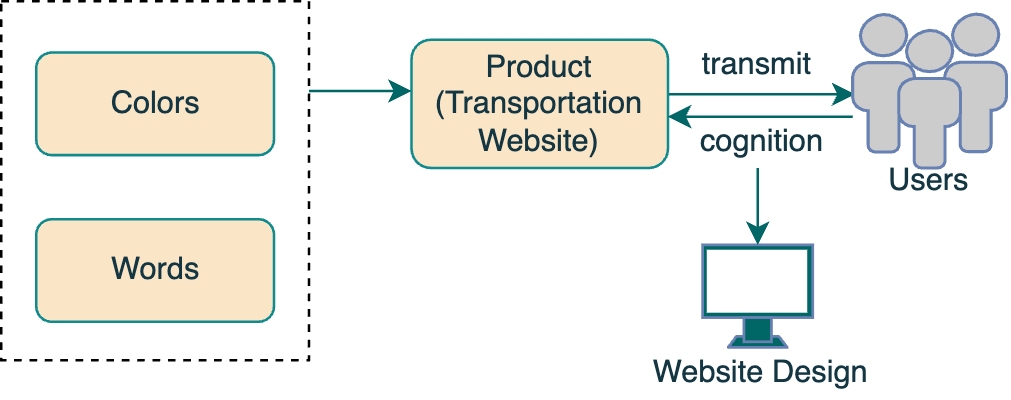}
    \caption{The formation process of web design}
    \label{kanseidesign}
\end{figure}

 In today's highly competitive market, product design and development success depends on functional attributes and producing desired emotional responses from consumers \cite{Lokman2010}. A key area of research involves studying user interaction with product interfaces to gain insights into user psychology and behavior\cite{Chen2013}. Figure \ref{kanseiidea} illustrates the general idea of the kansei engineering framework applied to website design. Figure \ref{kanseidesign} presents the formation process of web design.


Aesthetic, emotional, and other sensory aspects are just as significant as technical and objective requirements—some of which are challenging or impossible to articulate objectively \cite{pleasure}. In the practice of design, the designer must strike a balance between functional technology and emotional expressiveness, as well as between objective and subjective qualities.KE uses engineering technologies to investigate the connection between a product's design features and the user's emotions, optimizing the product design process \cite{appsc}. Integrating these aspects into product creation involves shifting the focus from functional to emotive needs of customers \cite{2008}.

Various measurement methods have been created to analyze physiological responses (e.g., heart rate, EMG, EEG), people's behaviors, factual and physical expressions, and spoken words \cite{new1}. Research on KE often focuses on the shape of industrial items, proposing a relationship between shape characteristics and affective information (11). These strategies simulate user emotive responses when manipulating product design characteristics \cite{mele}



A large body of studies published on KE shows that it has been widely applied in various industrial fields. As for website design, it was mainly used in areas where aesthetics and attractiveness play a pivotal role in consumer decision-making, e.g., in e-commerce for clothing and cosmetics. However, its application in sectors not traditionally focused on aesthetics, such as transportation, has been limited. This gap in research motivates us to explore the importance of emotional design in less aesthetic-focused fields.
The primary focus in the transportation industry has often been on functionality and efficiency. However, as this sector becomes more competitive, the importance of user experience is growing. 
By exploring these dimensions, we aim to explore how KE can be applied beyond traditional boundaries to improve user interaction and satisfaction in sectors not typically associated with aesthetic design. 


In this paper, we apply the Kansei approach, exploring the emotional
impact on the logistics website, using the site Reis.kz as the specific experimental context. This platform, being a marketplace for finding cargo and transport services, provides a suitable context for assessing the effectiveness of KE in improving user experience within a specific domain. The study identifies emotions associated with certain product features through a series of surveys, experiments, and evaluations. Afterward, these emotions and psychological feelings are converted into parameters for the appealing web design product.

The contributions of this study are as follows:
\begin{itemize}
    \item  The adapted Kansei-based model of web interface design, with the specific industrial case - Reis.kz platform.
    \item  Exploring the applicability of KE to a sector not traditionally linked with aesthetics—specifically, a transportation website—rather than industries focused on packaging-based products and websites strictly linked to aesthetics, e.g., clothing or beauty sites.
\end{itemize}
The paper is structured as follows. Section I is this introduction. Section II presents the overview of KE-based research works. Section III presents methods used in designing the Kansei-based web interface, including the survey design, statistical analysis, and others. The results of our experiment are presented in Section IV. Finally, Sections V, VI, and VII provide the Discussion, Conclusion, and ideas for future improvements, specifically for cargo and transport marketplace platforms.

\section{Related Work}

KE has been widely reported and extensively explored in the literature. KE is a methodology that focuses on designing innovative products by taking into account the emotions and preferences of customers  \cite{Nagamachi1999}. The aim is to create unique and unexplored products that cater to the specific needs and desires of consumers\cite{NAGAMACHI2010}. The concept was developed by M. Nagamachi from Hiroshima University. KE attempts to create unexplored products by considering and responding to consumers' emotions and preferences. Other research has indicated that designing emotional appeal products is more effective than focusing solely on usability\cite{Desmet2007}, \cite{Stappers2002}.

KE is a powerful tool successfully applied in various fields, such as designing packaging for herbal beverages\cite{Pratiwi2023} and designing beer can packaging to meet the emotional and sensitive needs of a specific audience\cite{Okamoto2017}. Similar research has been conducted that applies KE to car seat lever position \cite{car}. This study intended to establish the ideal lever position considering several constraints of vehicles, including safety, space, operability, and anthropometric variability, using ANOVA to analyze Kansei scores and CART together with Random forest to generate the rules. These studies have shown that Kansei Engineering can effectively determine the specific needs of the target customer for a particular product. It is also an effective technology that can translate these needs into the design domain.

KE can be applied when a product owner intends to develop a new product based on customers' perceptions\cite{LOKMAN2009}. Also, some works describe using KE to incorporate emotions into web design to improve user experience and satisfaction\cite{Howard2012}. The research\cite{Hussin2011} aims to compare the feelings of customers and product owners to determine if a website designed according to the guidelines of Kansei Web Design can produce the desired emotional responses. The idea suggests that when the customers' perceptions are closely aligned, it indicates the good achievement of the intended emotional experience for users. Another study proposes a data-driven method for user group-oriented Kansei evaluation \cite{Guo2021}.

Several studies applied KE for webpage design, including a job-hunting website homepage \cite{webpage1}, retailer website \cite{Mendoza2013}, implementing the design of online retail websites\cite{Lokman2007}, design of food E-commerce Mobile Application which was recently published \cite{mobilek}. A novel approach to building e-commerce applications allowing the fetching of items based on impressions like “romantic,” “formal,” or “elegant” and color perception was proposed in \cite{ieeeshop}. The research work \cite{Mendoza2013} has shown that web developers' design choices significantly impact how consumers perceive retailer websites. While traditional ergonomic recommendations prioritize functionality, recent studies have emphasized the importance of emotional engagement in website design to enhance attractiveness. Each product and service requires individual design solutions, with aesthetic design adding value by fostering emotional connections. Users increasingly seek emotional engagement in website interfaces, making their initial impressions of a website crucial. Website aesthetics, particularly color choices, play a key role in influencing consumers' first impressions of e-commerce websites. The study highlights the significant role of KE methodology in website design, effectively addressing users' need for "pleasure" in their online experience.

Several studies applied the Kansei approach to mobile design \cite{mobilek}, \cite{Yun2003}. The study \cite{Yun2003} aimed to measure user satisfaction with mobile phone designs using KE. A survey of 78 participants rated 50 mobile phone designs based on perceived image/impression attributes such as luxury, simplicity, beauty, color, texture, delicacy, harmony, salience, rigidity, and overall pleasure.

Information gathered from online sources has become crucial in various fields. Online retailing, in particular, provides plenty of customer feedback, making it essential to understand consumer sentiment towards products\cite{Liu2023}. The work \cite{Kandambi2022} underscores the importance of emotional responses and the influence of colors on consumer behavior, particularly in online contexts. This study explores how color concepts on designerwear clothing e-commerce websites impact user emotions and behaviors, utilizing Kansei Engineering. The study seeks to enhance emotional satisfaction and inform future website design decisions by validating outcomes and providing insights into effective color strategies.

A number of text mining approaches to KE have been proposed \cite{summar}, \cite{Liu2023}, \cite{sightseeing}, including
online product reviews based KE method
\cite{Liu2023}, extracting and summarizing emotional features from internet product descriptions and consumer reviews from Amazon \cite{summar}, developing a website for sightseeing travel using tweets analysis \cite{sightseeing}.

Another paper considers the application of KE in designing the package of bathtub salt, involving the analysis of 23 sample products and 30 adult subjects in the evaluation experiment \cite{inbook}.  Novel Kansei decision tree method was introduced and tested using the case of watch selection \cite{Hiroko1}. Another study introduces the Kansei-based bottle design system that includes rules and limitations for geometry, materials, manufacturing processes, top load, and interior pressure resistance. The focus is on a predicted case-based procedure that utilizes Artificial Neural Networks' \cite{mele}. The KE applied to the design of herbal beverage packaging has been considered by recent work \cite{Pratiwi2023}.

There are cases when research may face limitations due to its focus on the age of consumers \cite{Hussin2011}, for instance, aged 15-30, 30-45, or even older). Additionally, participants were from a local group rather than a worldwide population, suggesting that Kansei responses may vary based on racial features across different communities. Moreover, online retailing trends in a particular country could impact the research findings. If possible, a more diverse sample with varying characteristics would have yielded valuable insights for the new design\cite{Howard2012}.

From the literature review, many studies have shown that KE performs effectively with tangible products. As for web design, KE was applied in areas where attractiveness is highly important. Today's question is whether KE applies to platforms not typically associated with aesthetic design, e.g., logistics websites.

\section{Methods}
\subsection{Kansei Engineering Flow}
In this paper, we use Kansei Engineering type I, which has the following flow \cite{inbook}, \cite{Nagamachi2008ASS}: 
\begin{enumerate}
    \item Decision of a company strategy
    \item Collection of Kansei Words
    \item Setting of the Semantic Differential  scale
    \item Collection of product samples
    \item List of Item/Category
    \item Evaluation Experiment
    \item Statistical Analysis
    \item Interpretation of the analyzed data
    \item The explanation of data
    \item Collaboration with designers
\end{enumerate}

\begin{figure}[h]
    \centering
    \includegraphics[width=0.5\textwidth]{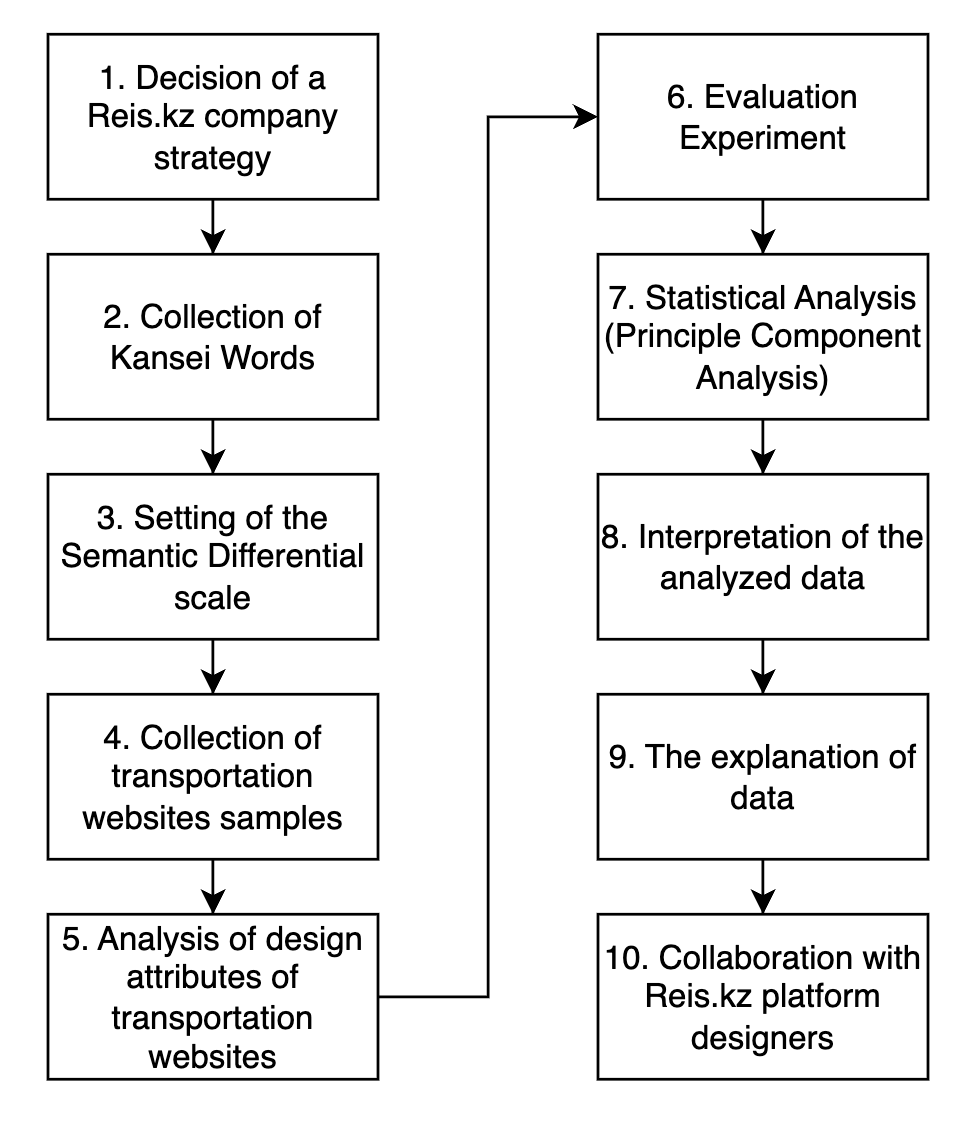}
    \caption{Kansei Engineering Flow}
    \label{fig:flow}
\end{figure}

The KE flow adapted to our context is presented in Fig. \ref{fig:flow}.
 \subsection{Decision of a company strategy}
   \begin{figure}[h]
    \centering
    \includegraphics[width=0.8\textwidth]{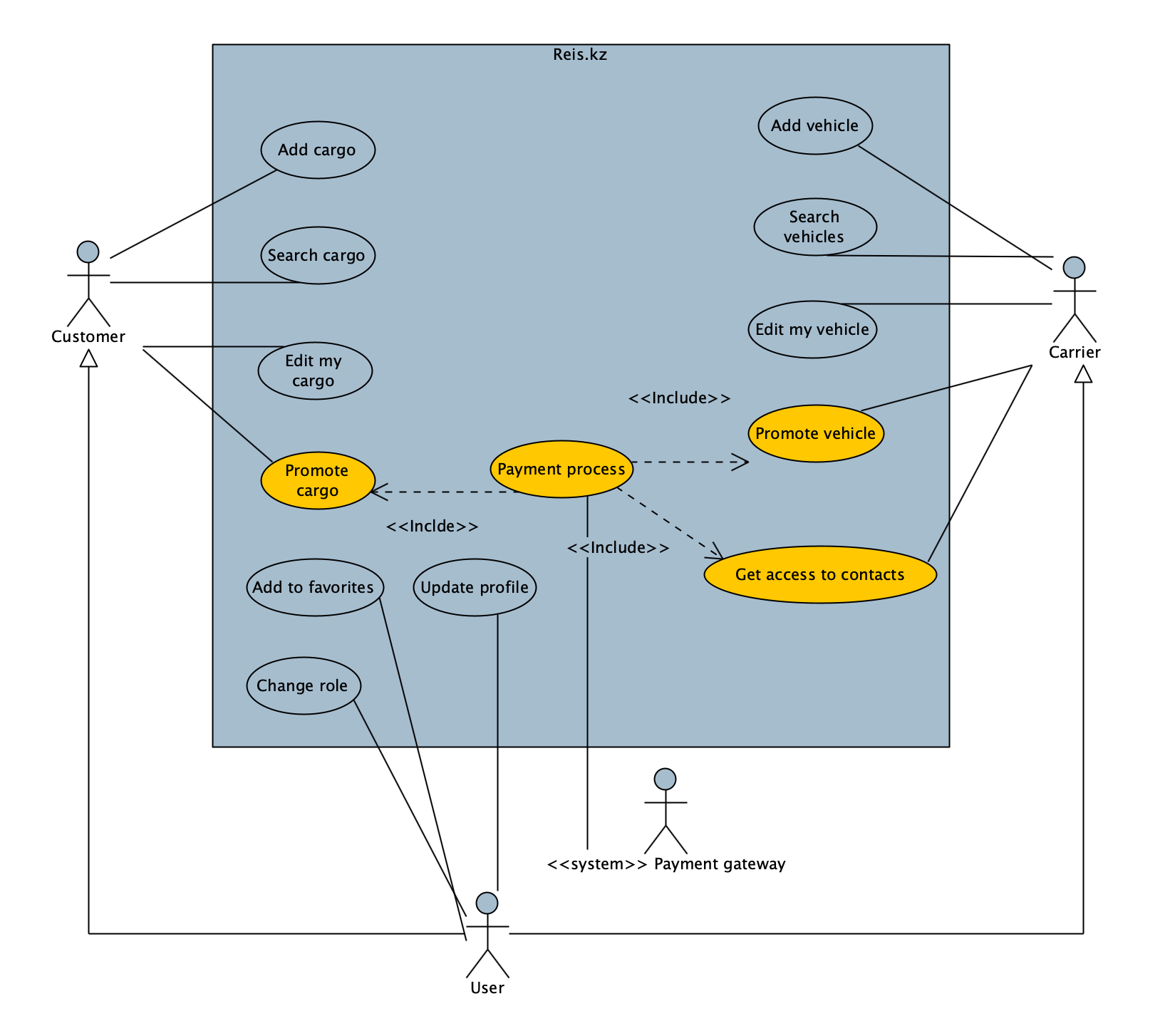}
    \caption{Use Case Diagram of considered Logistics Web application}
    \label{usecasefig}
\end{figure}

\begin{figure}[h]
    \centering
    \includegraphics[width=\textwidth]{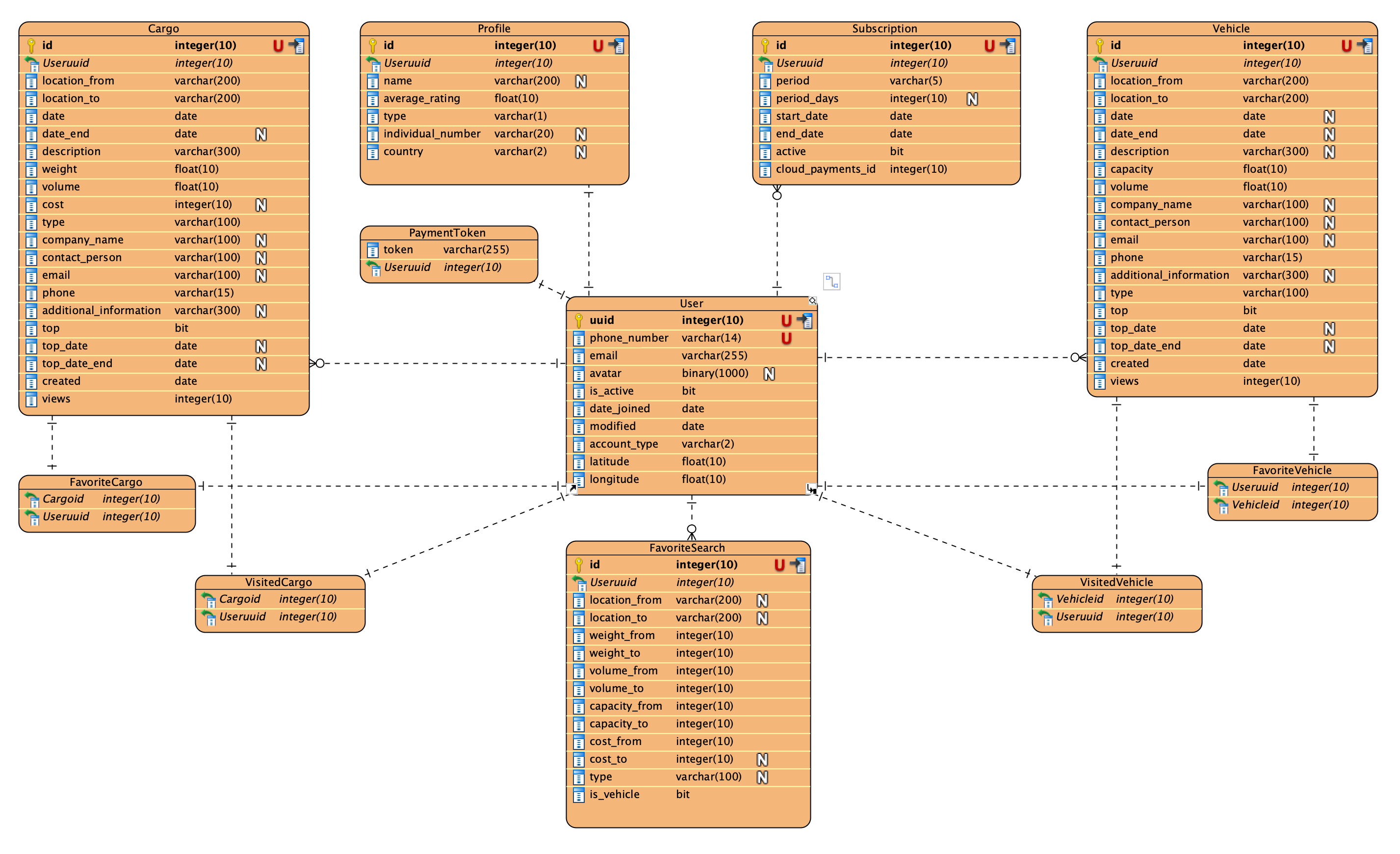}
    \caption{Entity Relationship Diagram of the Logistics Apllication}
    \label{erdfig}
\end{figure}

The Kansei engineer has to use the company strategy while applying Kansei Engineering to the new company product. We met with the representative of Reis.kz to investigate the company strategy. Use Case (see Fig. \ref{usecasefig}) and Entity Relationship (see Fig. \ref{erdfig}) diagrams were provided to us to understand the website functionally better. 

Reis.kz is a young company that aims to simplify and accelerate the process of international vehicle freight transport by connecting cargo owners with transportation providers through an online platform. The company's strategy focuses on creating a user-friendly, technologically advanced online platform that ensures ease of use and efficiency. Additionally, it aims to build trust and expand market reach through rigorous quality controls, intelligent matching algorithms, and strategic marketing efforts.

\subsection{Collection of Kansei Words}
     
The next step is to collect Kansei words related to the new product aligned with the strategy discussed in the previous step. A Kansei word is a term that describes the product context. These words can be adjectives or other grammatical types. To gain a thorough selection of words, the following sources are used \cite{new1}:
\begin{itemize}
    \item Pertinent Literature
 \item Experts
 \item Experienced Users
 \item Relating Kansei Studies
 \item Ideas, visions 
\end{itemize}
Some Kansei words were provided by Reis marketing experts and users, and some were adapted from related Kansei studies. As a result of our data collection process, the following Kansei words were collected (see Table \ref{kwrds}). Next, we grouped similar kansei words, choosing one representative kansei word for each group (see Table \ref{tab:kansei_words}). After reviewing research publications on evaluation metrics of emotional responses to online interfaces, products, and settings, a list of fifteen paired bipolar keywords was developed. 

\begin{table}[h]
\centering
\caption{Collected Kansei words}\label{kwrds}%
\begin{tabular}{@{}llll@{}}
\toprule
Kansei Word & Source & Kansei Word & Source \\
\midrule
Reliable &   Experts & Elegant &   Relating Kansei Studies \cite{Howard2012} \\
Efficient &   Experts & Bright &   Relating Kansei Studies \cite{Howard2012} \\
Secure &   Experts &Dynamic &   Relating Kansei Studies \cite{Howard2012}  \\
Flexible &   Experts &Beautiful &   Relating Kansei Studies \cite{Howard2012}  \\
Transparent &   Experts & Creative &   Relating Kansei Studies \cite{Howard2012} \\
Sustainable &   Experts & Elegant &  Relating Kansei Studies \cite{Howard2012} \\
Cost-effective & Experts & Well-structured  & Relating Kansei Studies \cite{Howard2012}  \\
Comprehensive &   Experts & Clear & Relating Kansei Studies \cite{Howard2012}  \\
Accessible &   Relating Kansei Studies \cite{mobilek} & Dynamic & Relating Kansei Studies \cite{discussion3}\\
Innovative &   Experts & Modern &  Relating Kansei Studies \cite{Howard2012}\\
Speedy &  Experienced Users & Informative &   Relating Kansei Studies \cite{Pratiwi2023}   \\
Streamlined & Experienced Users & Convenient &   Experienced Users \\
Scalable &   Experienced Users  & Accurate &   Experienced Users \\
Professional   & Experienced Users &Global &   Experienced Users  \\
Responsive &  Experienced Users &Integrated & Experienced Users  \\

\botrule
\end{tabular}
\end{table}

\begin{table}[h]
\centering
\caption{Kansei Words Grouping}\label{tab:kansei_words}%
\begin{tabular}{@{}lll@{}}
\toprule
Group & Kansei Words & Selected Word \\
\midrule
Reliability & Reliable, Secure, Sustainable & Reliable \\
Efficiency & Efficient, Speedy, Streamlined, Scalable & Efficient \\
Adaptability & Flexible, Accessible, Dynamic& Flexible \\
Transparency & Transparent, Clear & Transparent \\
Value & Cost-effective & Cost-effective \\
Depth & Comprehensive, Informative & Comprehensive \\
Innovation & Innovative, Creative, Modern, Dynamic & Innovative \\
Professionalism & Professional, Elegant & Professional \\
Responsiveness & Responsive & Responsive \\
Integration & Integrated & Integrated \\
Reach & Global & Global \\
Precision & Accurate & Accurate \\
Convenience & Convenient & Convenient \\
Structure & Well-structured & Well-structured \\
Aesthetics & Elegant, Beautiful, Bright & Elegant \\
\botrule
\end{tabular}
\end{table}

Finally, Kansei words were categorized based on Goldman's Model of Emotion \cite{emmodel}(see Table \ref{final}).
  \begin{table}[h]
\centering
\caption{Kansei Words Categorized According to Goldman's Model of Emotion}\label{tab:kansei_goldman_categories}%
\begin{tabular}{@{}llll@{}}
\toprule
Goldman's Categories & Selected Kansei Words & Bipolar Pairs &  \\
\midrule
Broadly Evaluative & Beautiful, Elegant & Beautiful - Not Beautiful &  \\
Novelty & Innovative & Innovative - Unoriginal &  \\
Complexity & Comprehensive & Comprehensive - Superficial &  \\
Intensity & Bright & Bright - Dull &  \\
Unity & Clear & Clear - Obscure &  \\
Prototypicality & Reliable & Reliable - Unreliable &  \\
\botrule
\label{final}
\end{tabular}
\end{table}

\subsection{Principal Component Analysis}
    Principal component analysis is one dimension reduction function commonly used to identify the most important factors that impact the user's emotions when using the program \cite{mobilek}.

\textbf{Covariance Matrix.} The covariance matrix \( \Sigma \) is calculated as follows:
\[
\Sigma = \frac{1}{n-1} (X - \overline{X})^T (X - \overline{X})
\]
Here, \( X \) is the data matrix, where each column represents a variable, and \( \overline{X} \) is the mean vector of the columns in \( X \).

\textbf{Eigenvalue Decomposition.} Eigenvalue decomposition is used to determine the eigenvectors and eigenvalues:
\[
\Sigma v = \lambda v
\]
where \( v \) are the eigenvectors and \( \lambda \) are the corresponding eigenvalues of the covariance matrix \( \Sigma \).

\textbf{Transformation to Principal Components.}
\[
T = X V
\]
where \( V \) is the matrix containing the eigenvectors, and \( T \) are the transformed principal components (scores).
    
\subsection{Survey Design}
  
  \subsubsection{Setting of the Semantic Differential  scale}
  \begin{figure}[h]
    \centering
    \includegraphics[width=0.5\linewidth]{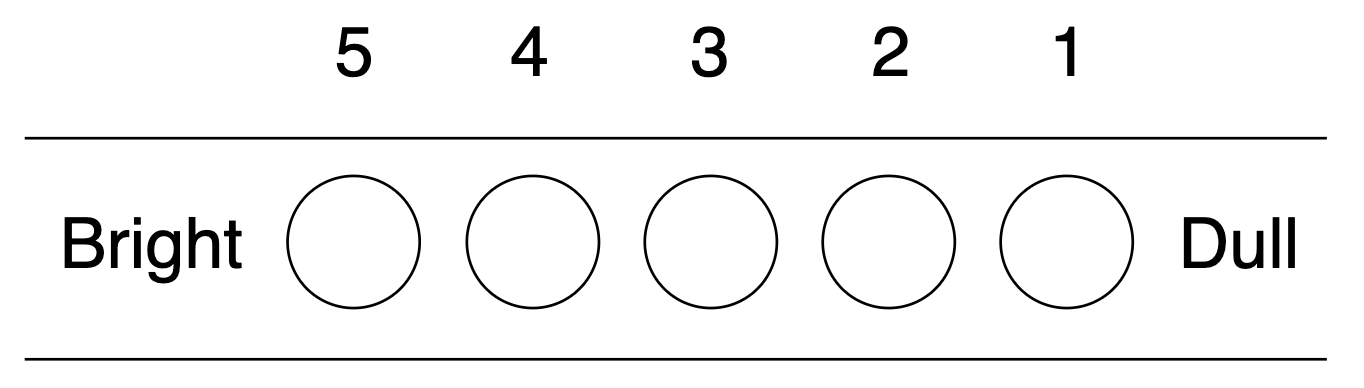}
    \caption{ Semantic Differential scale of Antithetical Adjectives}
    \label{scale}
\end{figure}
 The next step is to arrange the collected Kansei words on a 5-point Semantic differential (SD) questionnaire (Likert scale), where a positive Kansei word is represented by the greatest value and a negative Kansei word by the lowest value.  SD is a measurement tool using bipolar scales to assess an individual's subjective perception of and emotional responses to objects and ideas \cite{mobilek}. An example is provided in Fig. \ref{scale}.

\begin{figure}[h]
    \centering
    \includegraphics[width=0.9\linewidth]{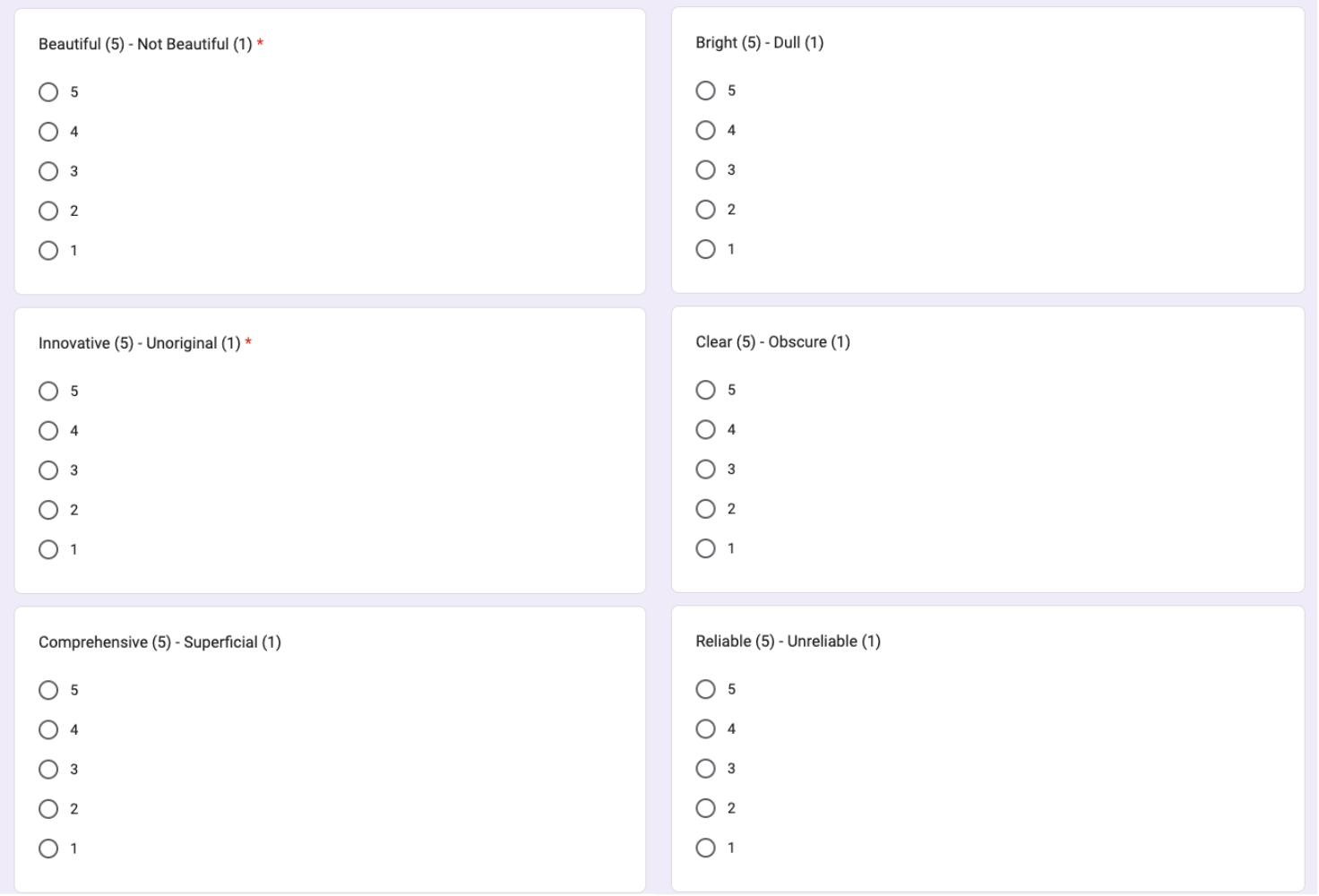}
    \caption{Survey design}
    \label{survey}
\end{figure}

 Fig. \ref{survey} presents the Google form used in the survey. As we see, there are five positions for evaluating each attribute. 

    \subsubsection{Collection of product samples}
\begin{figure}
        \centering
        \includegraphics[width=\linewidth]{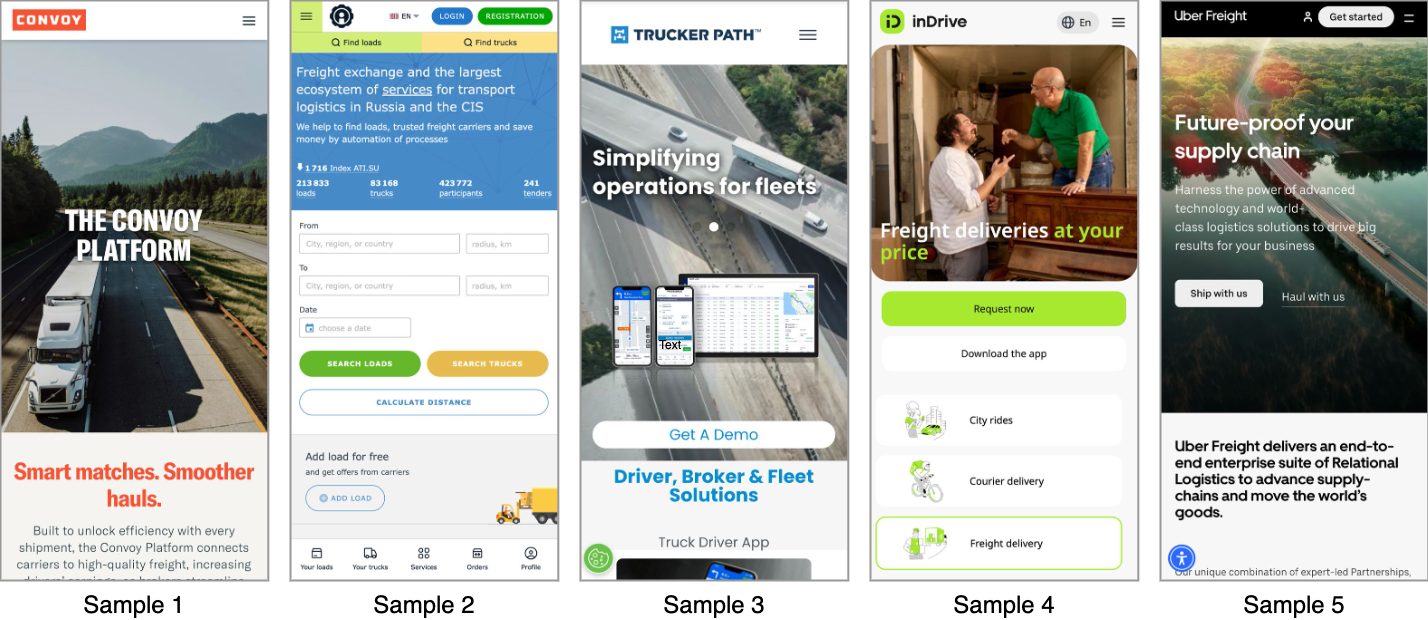}
        \caption{Samples of web designs of  transportation sites that have been used for data collection}
        \label{fig:samples}
    \end{figure}

  The first step was to select the experimental subjects and samples. Fig. \ref{fig:samples} showcases the website samples used in the evaluation experiment. The selected transportation companies include not only Kazakhstani ones but international ones as well. We also included benchmark companies in this list. 

\subsubsection{Survey Participants}

A group of participants with diverse backgrounds and varying experiences in using web interfaces were involved in the study. They were given questionnaires to provide feedback on emotional responses to the collected samples of transportation websites. Specifically, 30 participants (17 males and 13 females) were invited to evaluate the product samples by recording their emotions with Kansei words on the 5-point SD scale. The experiment was conducted using Google Forms.

    \subsection{List of Item/Category}
 This step implies the identification of the design features of the collected sample products, e.g., color, font, shape, size, and logo mark. We have selected six features from all products to analyze how each item influences emotions: 
\begin{itemize}
 \item Saturation and Intensity of the colors used on the page 
 \item Dominant color on the website 
 \item Number of colors used on the website 
 \item Brand logo visibility 
 \item Font size used on the page 
 \item Presence of images on the page and their proportion
\end{itemize}

This study analyzes the visual design elements used in various digital platforms. So, each platform's design is evaluated based on saturation, intensity, dominant colors, color count, logo visibility, font size, and image presence.

\subsection{Collection of Kansei colors}

In KE, color models play a crucial role by helping to understand the emotional responses colors evoke in users. By analyzing colors through models like RGB, CMYK, or Fuzzy color models \cite{ieeecolors}, designers can align the emotional impact of colors with user preferences.  This research focuses on the emotional impact of single Kansei colors rather than using multiple-color palettes. Nevertheless, color harmony \cite{harmony} can be utilized in KE to design web interfaces so that the multiple colors used are aesthetically cohesive and pleasing. 

We used several basic colors as mentioned in other research works \cite{Tharangie2009}, \cite{Tharangie2008}, \cite{bian2018}. In addition, we used colors according to characteristics obtained from the survey. Selected Kansei colors and a detailed explanation of their choice are presented in the Experimental Results section (Table \ref{tab:color_scheme} and the survey form is shown in Fig. \ref{fig:colorsurvey}).


\section{Experimental Results}
\subsection{Evaluation Experiment}
After identifying Kansei factors and conducting a survey, we need to translate them into design elements.
The design properties of selected samples were described according to the specifications described above (see Table \ref{props_descr}).
 \begin{table}[h]
\caption{Design Attributes of Transportation Websites}\label{tab:website_comparison}
\begin{tabularx}{\textwidth}{@{}X*{5}{>{\centering\arraybackslash}X}@{}}
\toprule
\textbf{Attribute} & \textbf{Sample 1} & \textbf{Sample 2} & \textbf{Sample 3} & \textbf{Sample 4} & \textbf{Sample 5} \\
\midrule
\textbf{Saturation, Intensity} & High saturation with vivid green and blue colors & Moderate, clear yet engaging. & High, with vivid imagery, cool colors & Moderate to high with warm tones & High, with vivid imagery\\
\midrule
\textbf{Dominant Color} & Orange, green and blue (landscape and sky) & Green and blue, signaling professionalism. & Blue and gray & Green and earth tones & Green and brown (natural tones) \\
\midrule
\textbf{Color Count} & Limited, focused on a few strong colors & Moderate, clean with green, blue, and gray. & Limited, with a professional look & High, with multiple color blocks & High, diverse range \\
\midrule
\textbf{Logo Visibility} & Clearly visible at the top & Prominent at the top left. & In the header & In the top header & In the header \\
\midrule
\textbf{Font Size }& Large and readable & Varied for headings and body text, enhancing readability. & Large, clear & Varied, large buttons, smaller descriptive text & Large and impactful \\
\midrule
\textbf{Image Presence }& Major element, background & Minimal, focusing on text and interactive elements. & Significant, upper half & Substantial, significant portion & Dominant, scenic background \\
\botrule
\label{props_descr}
\end{tabularx}
\end{table}

   The next step is to analyze the evaluated data using statistical methods to obtain the relations between each Kansei word and product samples. The Principal Component Analysis (PCA) was performed on the survey dataset, which is a very useful tool for decision-making of the new product
strategy. Here's what has been found:

    \begin{figure}[h]
        \centering
        \includegraphics[width=0.8\linewidth]{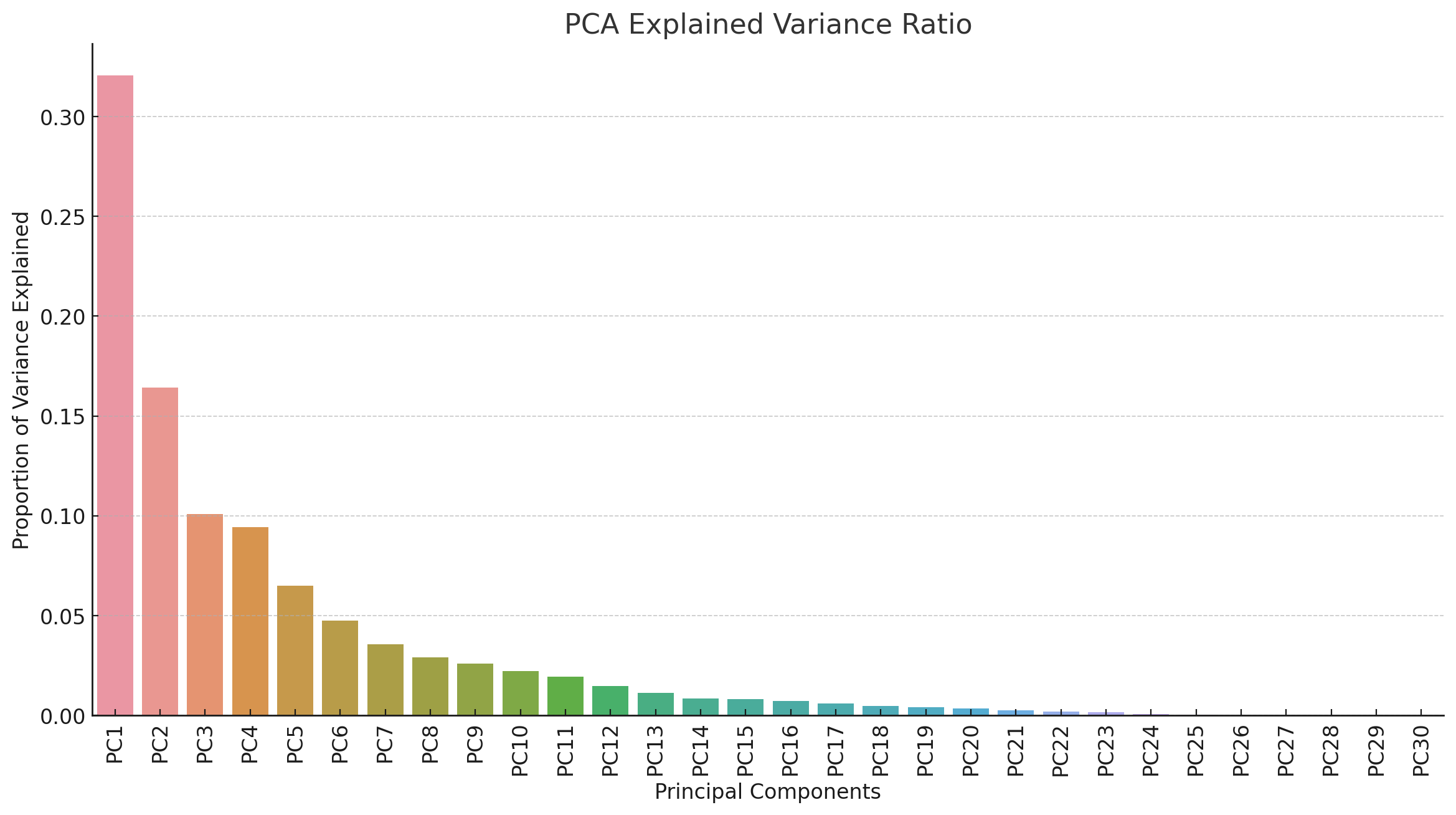}
        \caption{Explained Variance Ratio}
        \label{fig:var}
    \end{figure}

    \begin{figure}[h]
        \centering
        \includegraphics[width=0.8\linewidth]{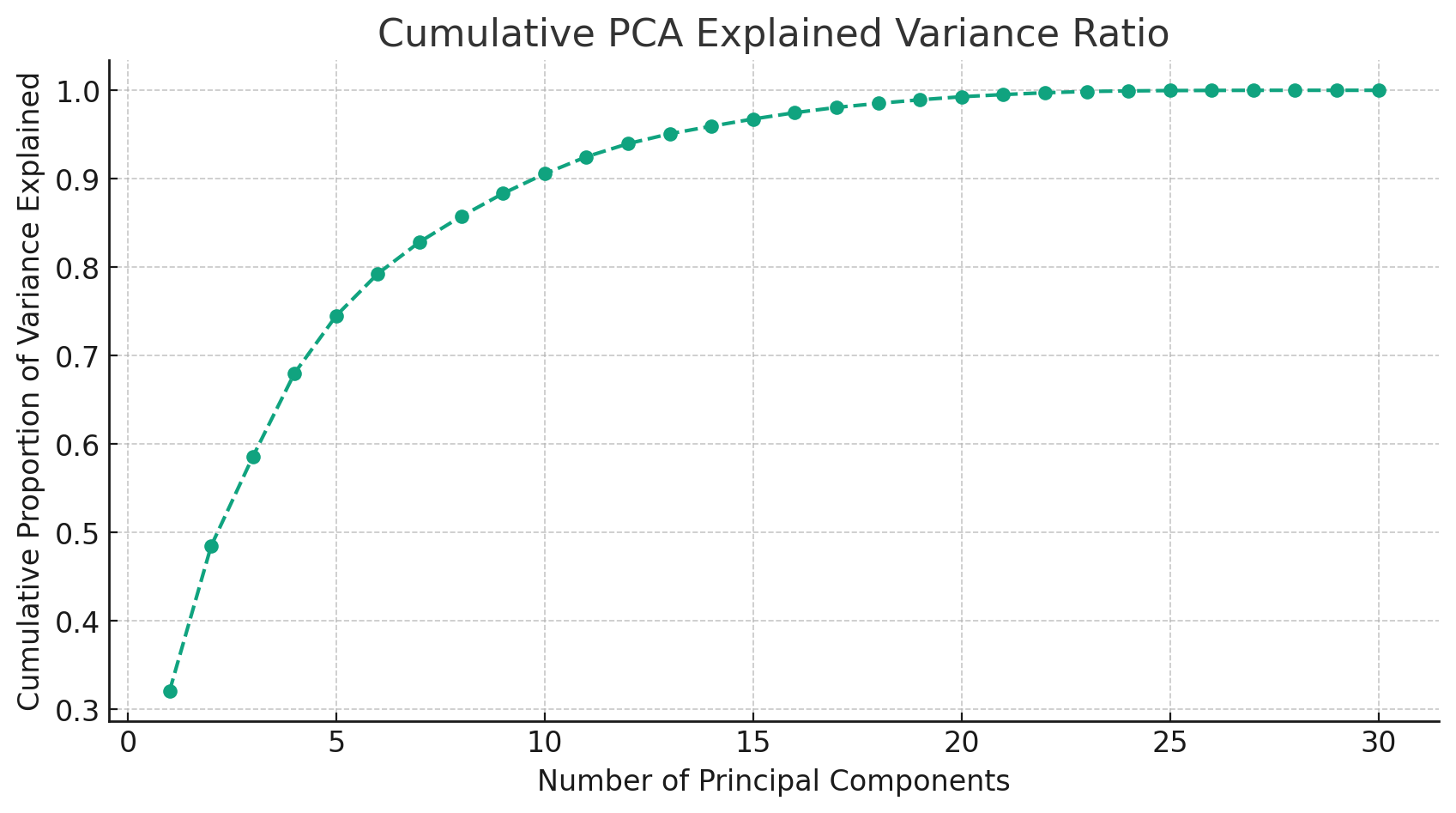}
        \caption{Cumulative Variance}
        \label{fig:cum}
    \end{figure}

\begin{itemize}
    \item \textit{Explained Variance Ratio}. The first few principal components capture a significant amount of the total variance in the data. The first principal component (PC1) alone accounts for approximately 32\% of the variance, and the second (PC2) accounts for about 16\%. The first two principal components account for nearly 48\% of the total variance (see Fig. \ref{fig:var}).
    \item \textit{Cumulative Variance.} The cumulative variance explained by the principal components indicates that around 68\% of the total variance is explained by the first four components, and about 99\% of the variance is explained by the first twenty-four components (see Fig. \ref{fig:cum}).
\end{itemize}

Next, we create a biplot, a type of plot that can be used to visualize the results of PCA. It shows both the scores (the transformed PCs values for each observation) and the loadings (the coefficients that describe the linear combination of the original variables to get the components) on the same plot. The biplot helps us understand how the original variables contribute to the PCs and how the observations are distributed according to these components.

\begin{figure}[h]
        \centering
        \includegraphics[width=0.7\linewidth]{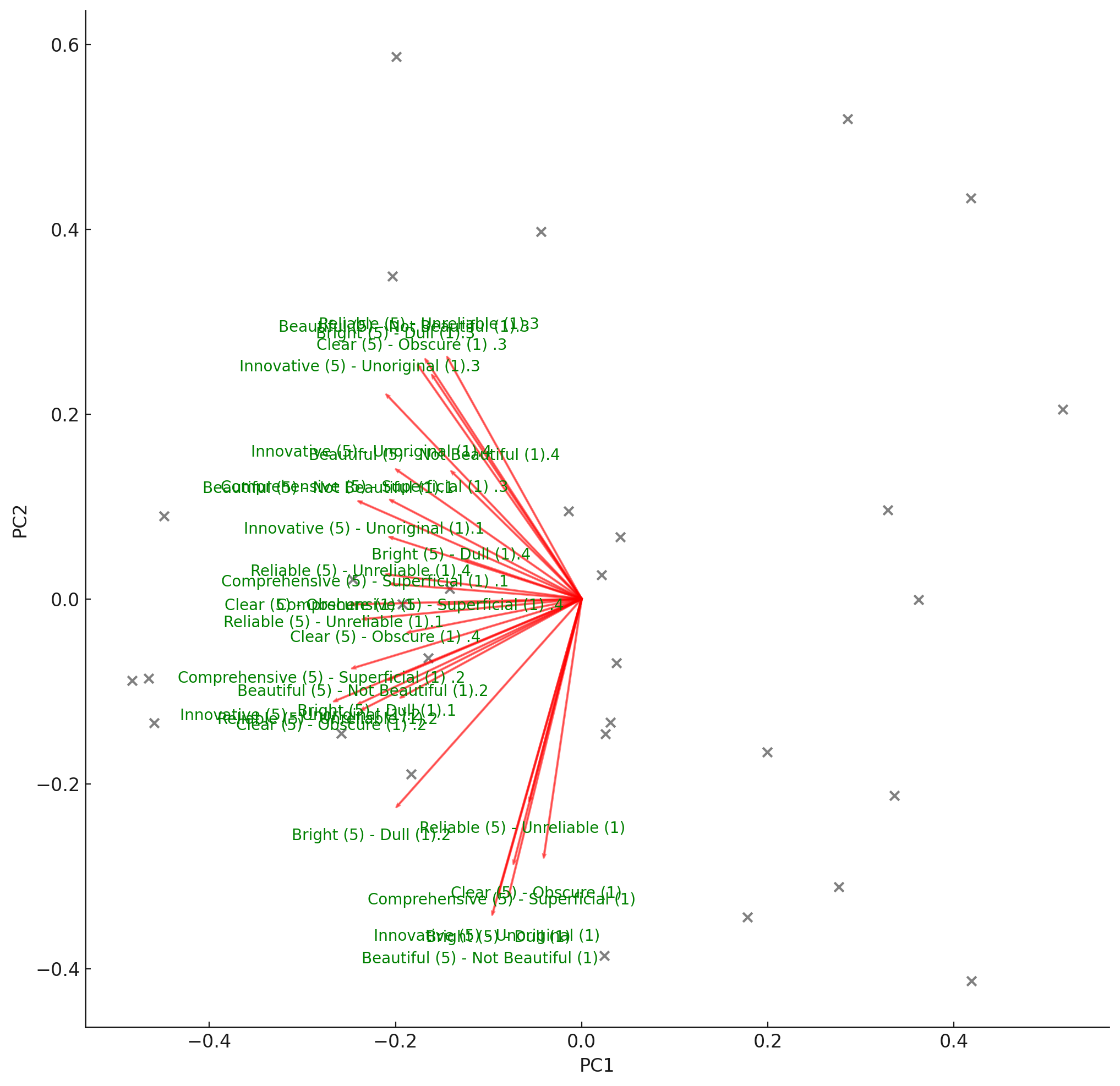}
        \caption{PCA biplot}
        \label{fig:pca}
    \end{figure}

Let's visualize the first two principal components (PC1, PC2) in a biplot. The result is shown in Fig. \ref{fig:pca}. The grey points represent the observations (respondents) regarding their PC1 and PC2 scores. The red arrows indicate the direction and magnitude of the original variables in relation to the principal components, allowing us to see which variables contribute most to the differences between observations.

The longer the arrow, the more that variable contributes to the component. Arrows pointing in the same direction indicate that the variables are positively correlated, while arrows pointing in opposite directions indicate negative correlations.

    \subsection{Interpretation of the analyzed data}
The main goal of this step is to find the relationship between Kansei factors and design specifications using the data collected.



\begin{figure}
        \centering
        \includegraphics[width=0.8\linewidth]{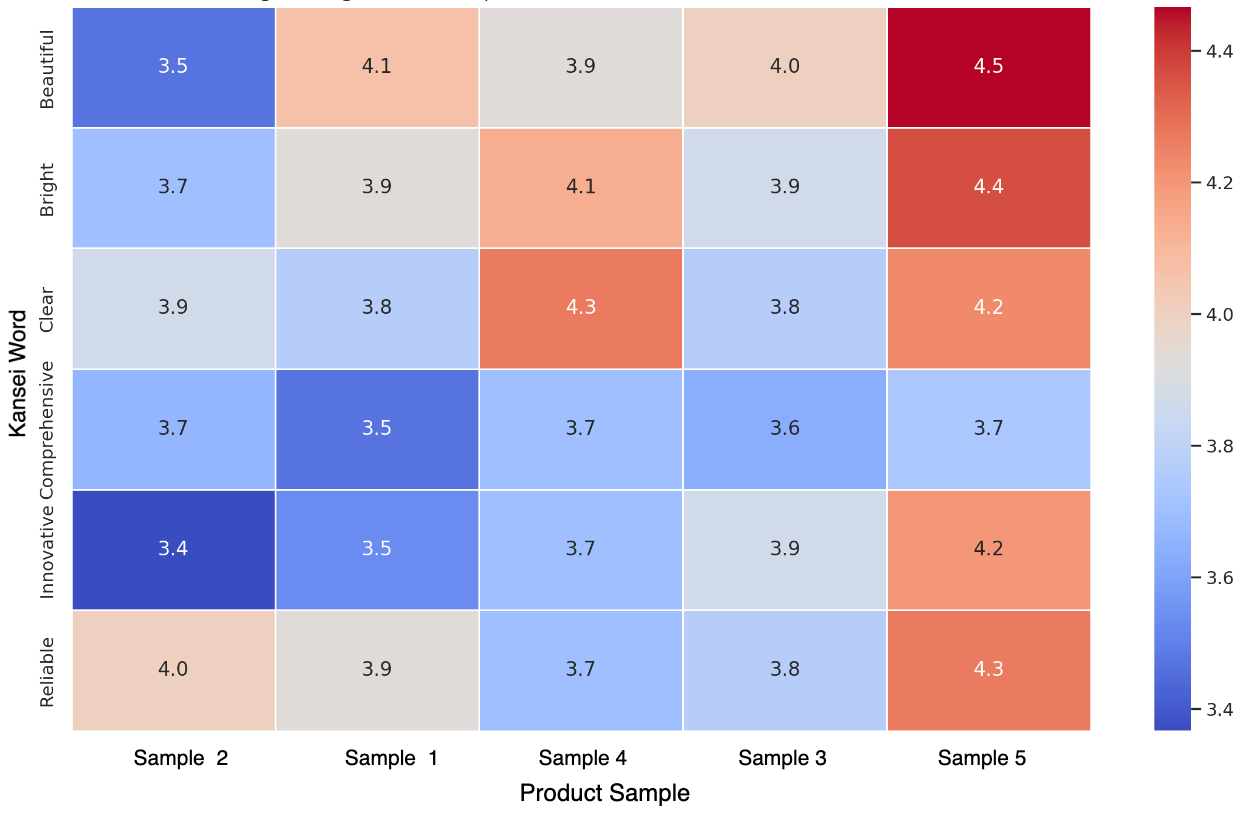}
        \caption{Average Ratings across product samples and Kansei Words}
        \label{fig:heatmap}
    \end{figure}

\begin{figure}
        \centering
        \includegraphics[width=\linewidth]{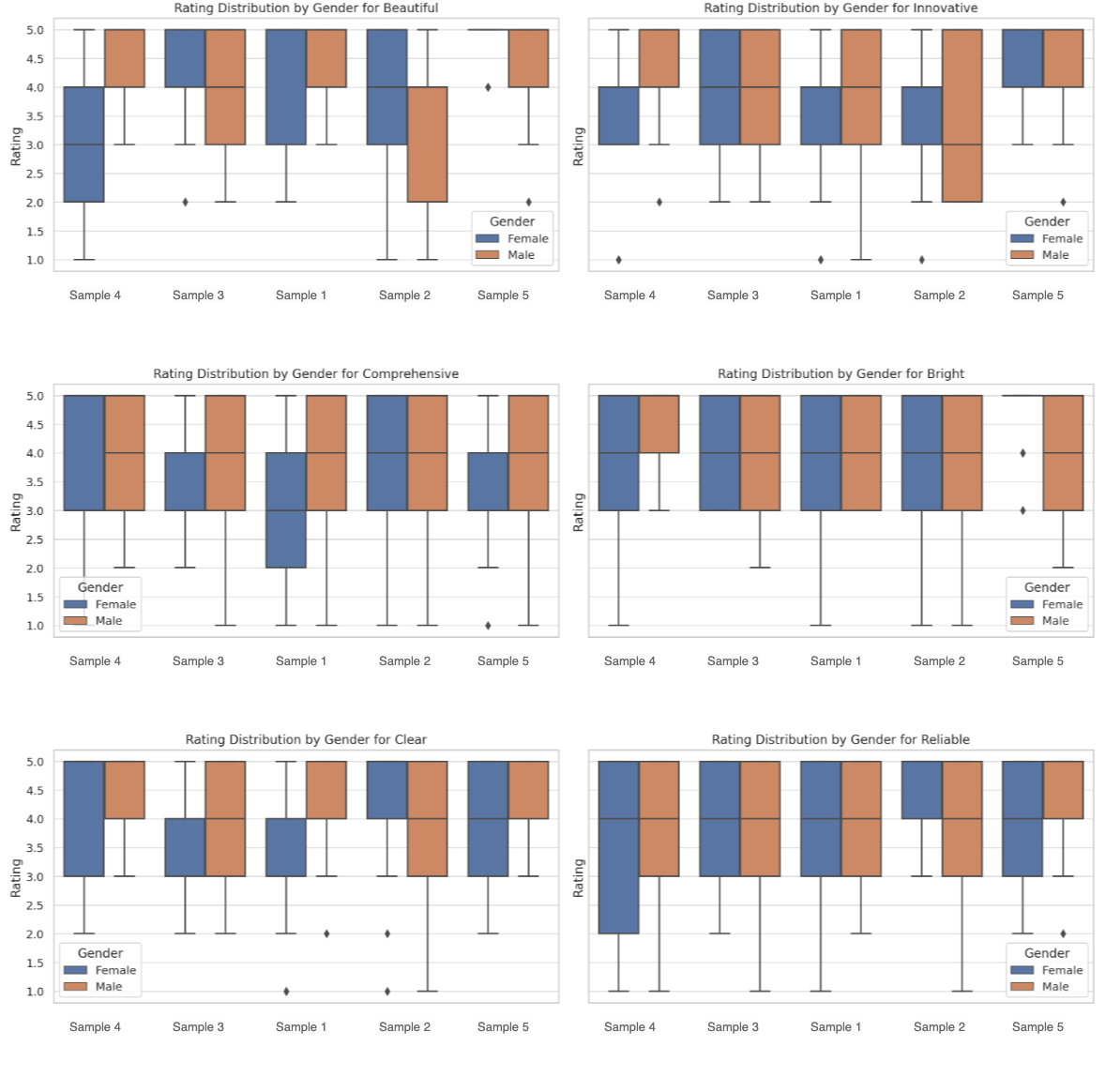}
        \caption{The distribution of ratings for each emotion by gender across the different product samples}
        \label{fig:boxplots}
    \end{figure}

Fig. \ref{fig:heatmap} presents a clear overview of how each product sample is rated across different emotional aspects (Kansei words) without differentiation by gender. Fig. \ref{fig:boxplots} illustrates the distribution of ratings for each emotion by gender across the different product samples. You can observe the median, interquartile range, and any potential outliers. For most emotions, the ratings are similar between males and females, suggesting a general agreement. However, the Kansei word "Beautiful" stands out as an exception. Here, the ratings significantly differ between genders, indicating that males and females have different views on what they consider beautiful. This shows that beauty is seen differently by each gender.

    The PCA suggests that distinct underlying factors influence the perception of logistics design. The first principal component (PC1), which accounts for the largest variance, is significantly influenced by attributes such as \textit{"Beautiful," "Clear," }and \textit{"Innovative."} This PC will be named "Elegant Innovation". This indicates that aesthetics and clarity of design, coupled with the perception of innovation, are principal factors that contribute to the overall impression of a design in logistics. 
    
    The second principal component (PC2) appears to be influenced by\textit{ "Comprehensive," "Reliable," }and \textit{"Bright" }attributes. We can name this PC as "Radiant Trust". For instance, products rated high on PC1 tend to be seen as beautiful, clear, and innovative, while those that score high on PC2 might be considered comprehensive, reliable, and bright.

When both PC1 and PC2 are considered, it is evident that an effective design in the logistics industry is not merely about functionality; instead, the interplay between these characteristics shapes the user’s emotional response.

We need to identify key design features associated with Kansei words, explain the data interpretation to the company designer(s), apply them to website design, and collaborate on creating the new design. 
    


Now let us perform matching of PCA-based strategic Kansei to design features. We have six features, and we need to select the ones that have a larger impact on Kansei. From our previous PCA analysis, respondents differentiate logistics designs based on two main dimensions: "Elegant Innovation" dimension composed of \textit{"Beautiful," "Clear,"}and \textit{"Innovative"}(PC1) and ”Radiant Trust" dimension (PC2) composed of \textit{ "Comprehensive," "Reliable," }and \textit{"Bright" } Kansei words:

\begin{itemize}
    \item \textbf{PC1.} To identify which features are leading in terms of the ratings for "Beautiful," "Clear," and "Innovative," we calculated the average ratings for these specific Kansei words and then compared them across the product samples leading in these criteria. Average ratings are as follows: Sample 5: 4.30, Sample 4 - 3.97, Sample 3 - 3.88. Based on the analysis of similar design patterns in this samples (see Table \ref{props_descr}), we selected two features, namely \textit{"Saturation and Intensity of the colors"}, value is \textit{"High"} (F1) and \textit{"Brand Logo visibility"}, value is \textit{"In the header"}(F2). 

    \item \textbf{PC2.} It is composed of "Comprehensive," "Reliable," and "Bright" kanseis. related to readability. We calculated the averages for these specific emotions and identified which companies are leading based on these criteria. Average ratings are Sample 5: 4.12, Sample 4 - 3.84, Sample 2 - 3.79. Analyzing comparable design patterns in these samples (see Table \ref{props_descr}), we identified two features, namely: \textit{"Color Count"} as \textit{"High"} (F3), and \textit{"Font Size"} is \textit{"Varied, large"} (F4).

\end{itemize}

Finally, we can integrate all items/categories related to Kansei of PC1 and PC2 and develop a new website design in collaboration with company representatives.

    \subsection{Data explanation and collaboration with designers}

   Based on the analysis of evaluated data, we identified the new design specifications.  Combining design criteria from KE implementation with the experience and skill of product designers can generate a successful Kansei product. But first, we need to identify the dominant colors based on specification \textit{High Saturation and Intensity}.

\begin{figure}[h]
    \centering
    \begin{minipage}[b]{0.29\textwidth}
        \centering
        \includegraphics[width=\textwidth]{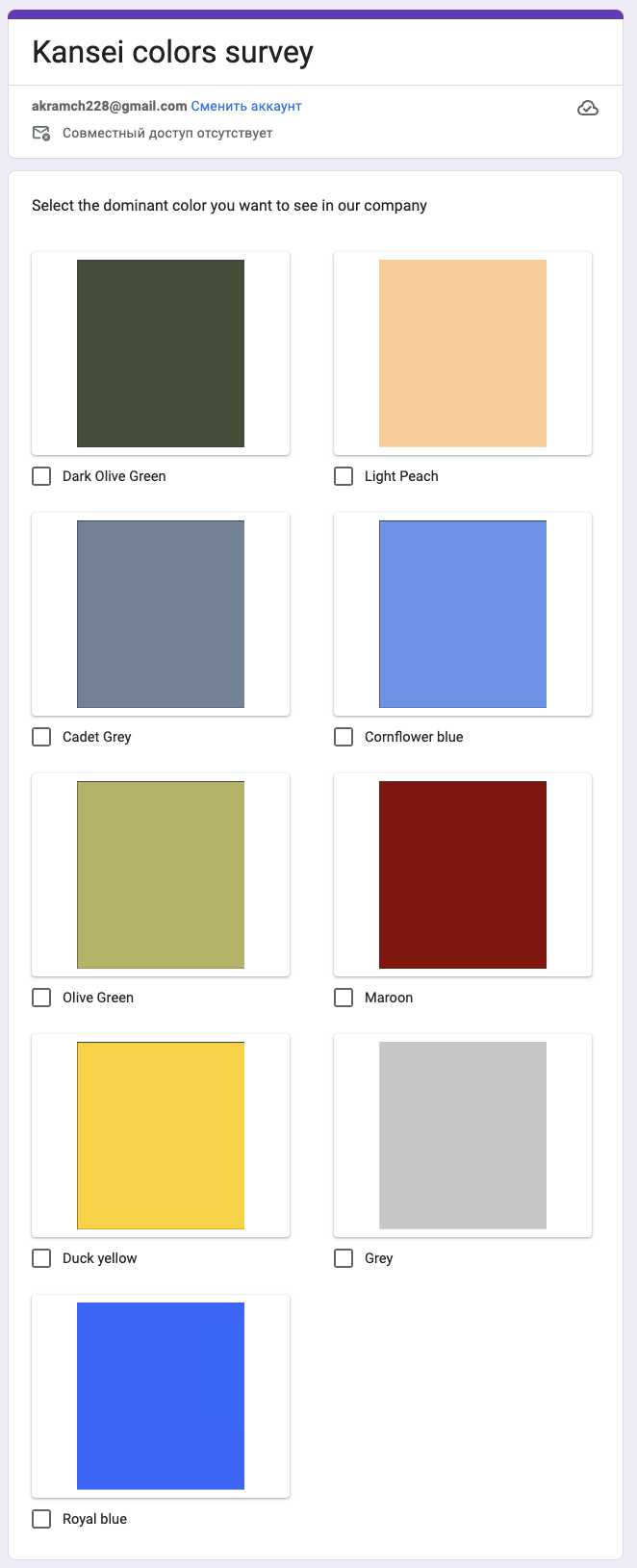}
        \caption{Colors survey}
        \label{fig:colorsurvey}
    \end{minipage}
    \hfill
    \begin{minipage}[b]{0.7\textwidth}
        \centering
        \includegraphics[width=\textwidth]{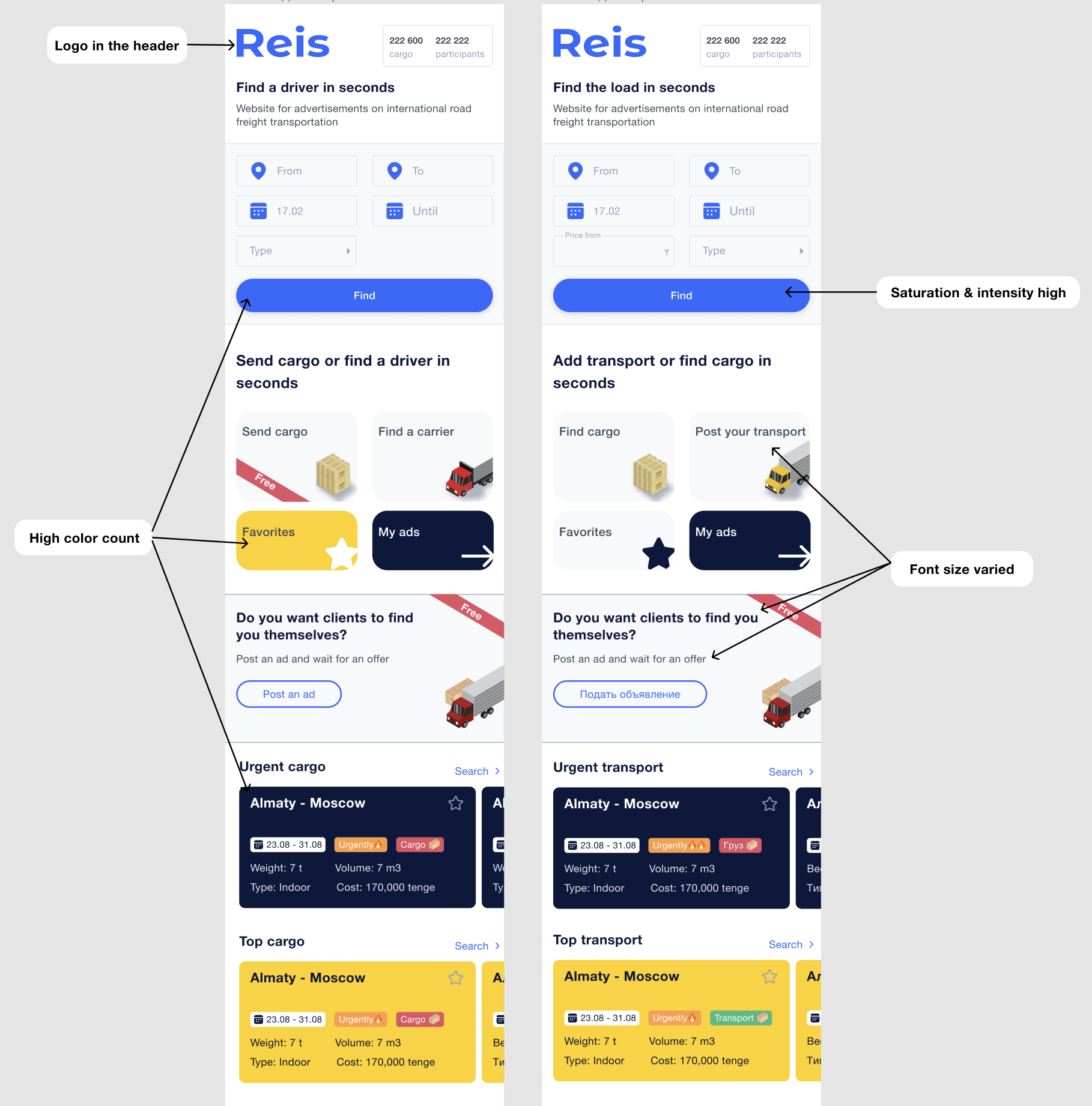}
        \caption{The new mobile website design}
        \label{finfig}
    \end{minipage}
\end{figure}

Selected Kansei colors are presented in Table \ref{tab:color_scheme}, and the survey form is shown in Fig. \ref{fig:colorsurvey}. This survey aimed to evaluate color preferences among design options in the context of KE. When evaluating different colors, the focus was on understanding which resonated most with the company vision and was of high saturation and intensity. Participants were ten individuals familiar with the Reis.kz platform, including company designers. The follow-up survey type was multiple choice, allowing them to select multiple colors they found appealing. The survey included nine color choices, as shown in Table \ref{tab:color_scheme}. Colors were collected from related studies, survey results incorporating product samples, or from one's ideas and vision. 

Table \ref{tab:color_scheme} shows that "Royal blue" got eight votes, indicating a strong preference among the experts. The colors "Grey" and "Duck Yellow" followed closely behind with seven and five votes. Colors like "Light Peach" and "Olive Green" received fewer votes, indicating a lower preference among the experts.

\begin{table*}[]
\centering
\caption{Selected Kansei colors}\label{tab:color_scheme}
\begin{tabular}{@{}ccccc>{\centering\arraybackslash}p{2.5cm}@{}}

\toprule
\textbf{Color} & \textbf{RGB} & \textbf{Name} & \textbf{Votes} & \textbf{Design} & \textbf{Source} \\
\midrule
\begin{tikzpicture}
    \definecolor{fillcolor}{rgb}{0.255, 0.31, 0.216}
    \fill[fillcolor] (0,0) rectangle (0.7,0.7);
\end{tikzpicture} & (65, 79, 55) & Dark Olive Green &  5 & \includegraphics[width=2cm]{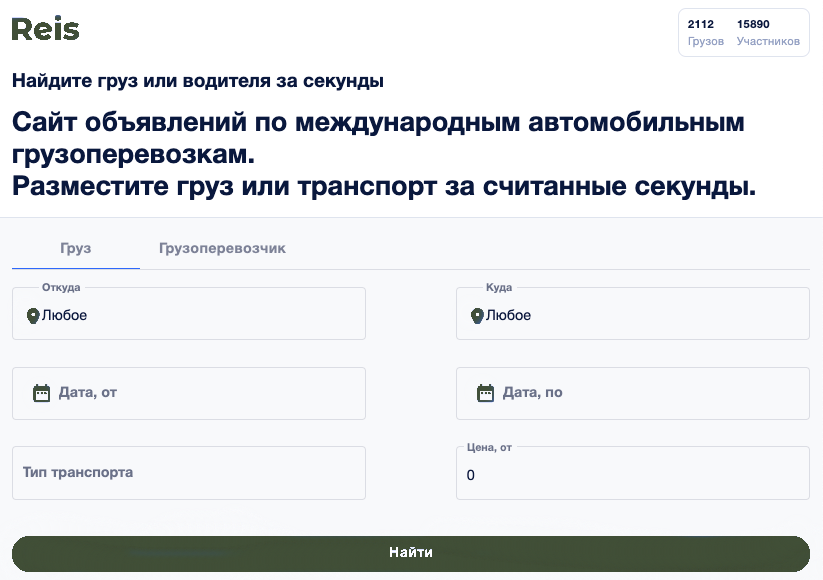} & Ideas, visions\\
\begin{tikzpicture}
    \definecolor{fillcolor}{rgb}{1, 0.796, 0.576}
    \fill[fillcolor] (0,0) rectangle (0.7,0.7);
\end{tikzpicture} & (255, 203, 147) & Light Peach &  3 & \includegraphics[width=2cm]{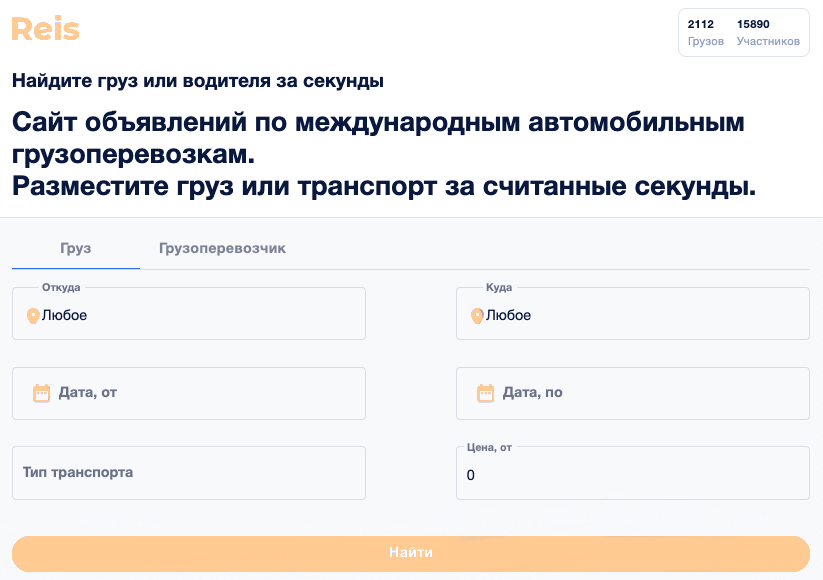} & Relating Kansei Studies\cite{Tharangie2008}\\
\begin{tikzpicture}
    \definecolor{fillcolor}{rgb}{0.435, 0.51, 0.596}
    \fill[fillcolor] (0,0) rectangle (0.7,0.7);
\end{tikzpicture} & (111, 130, 152) & Cadet Grey &  4 & \includegraphics[width=2cm]{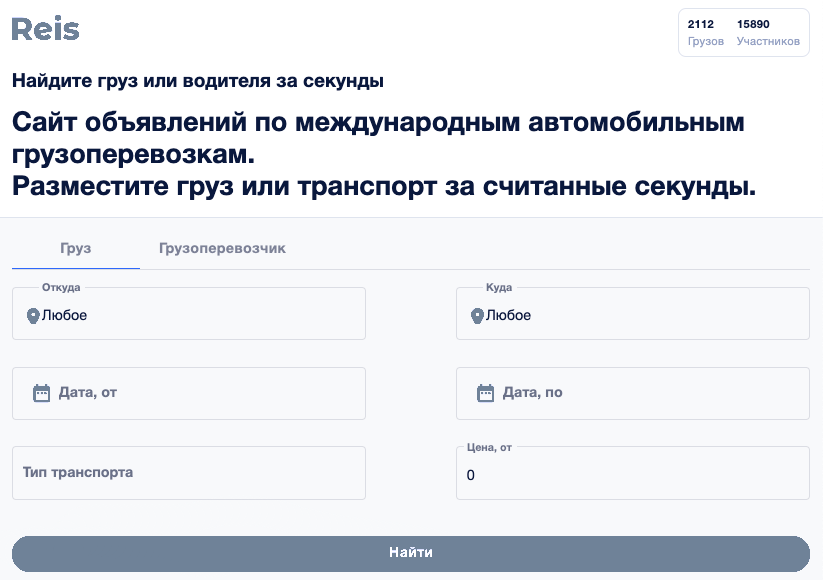} & Relating Kansei Studies\cite{bian2018} \\
\begin{tikzpicture}
    \definecolor{fillcolor}{rgb}{0.392, 0.584, 0.929}
    \fill[fillcolor] (0,0) rectangle (0.7,0.7);
\end{tikzpicture} & (100, 149, 237) & Cornflower Blue &  6 & \includegraphics[width=2cm]{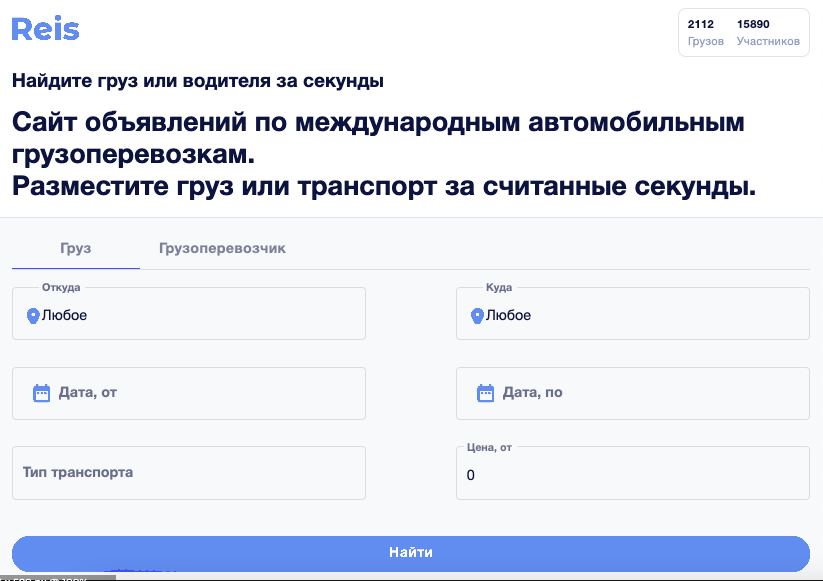} & Survey \\
\begin{tikzpicture}
    \definecolor{fillcolor}{rgb}{0.710, 0.702, 0.361}
    \fill[fillcolor] (0,0) rectangle (0.7,0.7);
\end{tikzpicture} & (181, 179, 92) & Olive Green &  2 & \includegraphics[width=2cm]{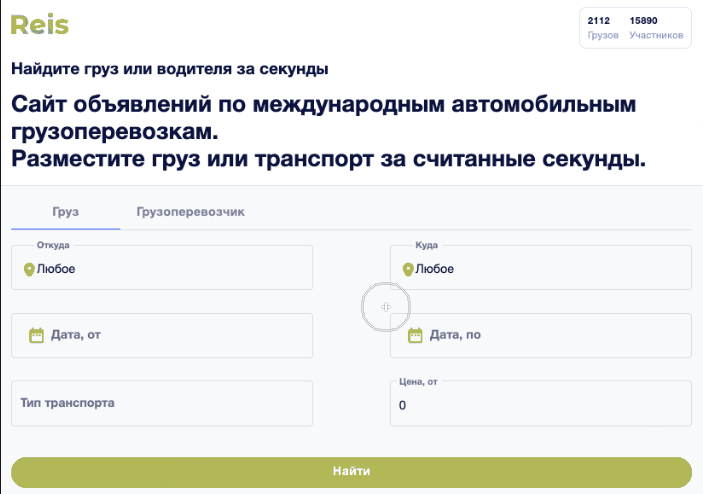} & Survey\\
\begin{tikzpicture}
    \definecolor{fillcolor}{rgb}{0.545,0,0}
    \fill[fillcolor] (0,0) rectangle (0.7,0.7);
\end{tikzpicture} & (139, 0, 0) & Maroon &  4 & \includegraphics[width=2cm]{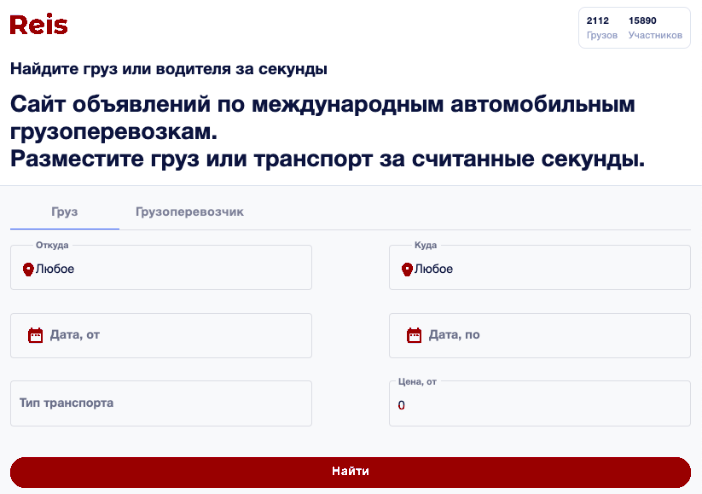} & Ideas, visions\\

\begin{tikzpicture}
    \definecolor{fillcolor}{rgb}{1, 0.816,0}
    \fill[fillcolor] (0,0) rectangle (0.7,0.7);
\end{tikzpicture} & (255, 208, 0) & Duck yellow & 5 & \includegraphics[width=2cm]{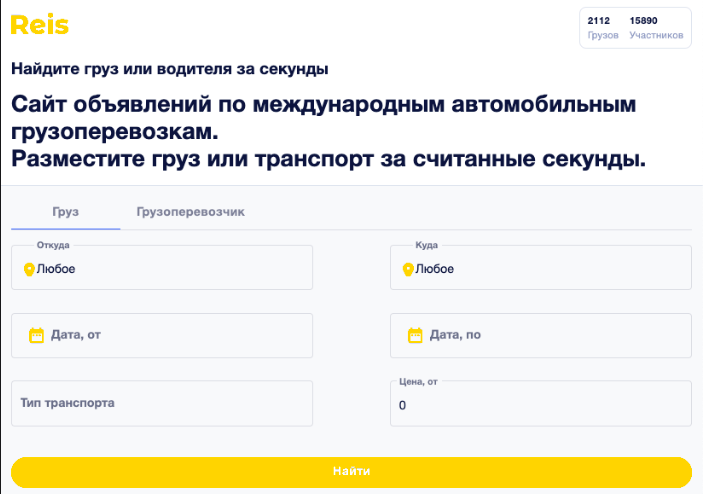} & Ideas, visions\\
\begin{tikzpicture}
    \definecolor{fillcolor}{rgb}{0.773, 0.776, 0.780}
    \fill[fillcolor] (0,0) rectangle (0.7,0.7);
\end{tikzpicture} & (197, 198, 199) & Grey &  7 & \includegraphics[width=2cm]{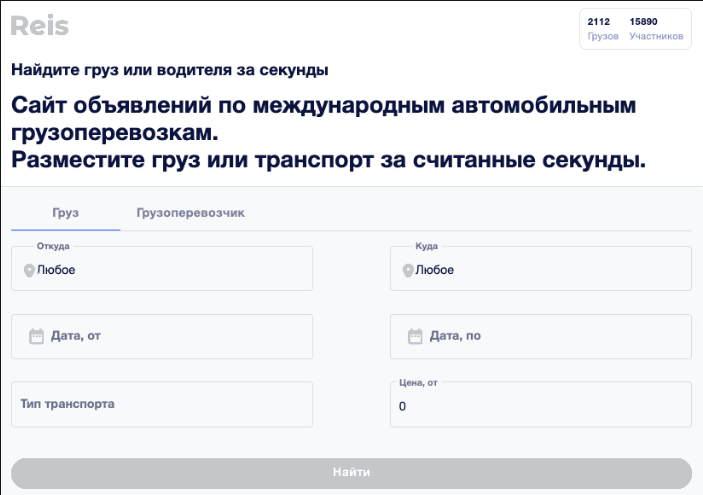} & Ideas, visions\\
\begin{tikzpicture}
    \definecolor{fillcolor}{rgb}{0.176, 0.408, 0.996}
    \fill[fillcolor] (0,0) rectangle (0.7,0.7);
\end{tikzpicture} & (45, 104, 254) & Royal blue &  8 & \includegraphics[width=2cm]{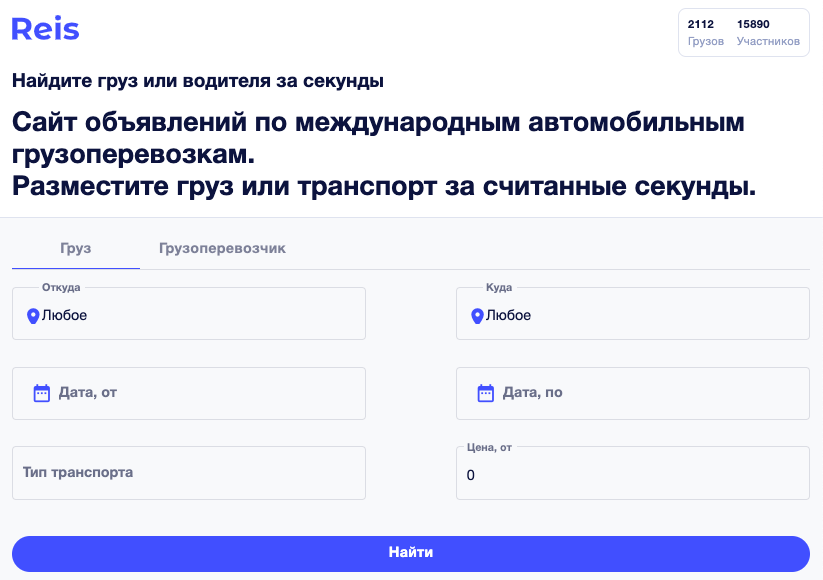} & Ideas, visions\\
\botrule
\end{tabular}
\end{table*}


The updated components in the design align with the PCA findings (see Fig. \ref{finfig}):
\begin{itemize}
    \item \textbf{F1:} \textit{"Saturation and Intensity of the colors is High"}. This was well reflected in the updated design, which uses Royal blue, Duck Yellow, Grey, and Black colors with appropriate saturation to enhance visual appeal. 
    \item \textbf{F2:} \textit{"Brand Logo visibility is in the header"}. In the updated design, the logo is well visible and placed in the header to enhance brand recognition. 
     \item \textbf{F3:}  \textit{"Color Count is High"}. We used four carefully selected colors, considering other specifications and surveys.
      \item \textbf{F4:}  \textit{"Font Size is Varied, Large"}. We optimized font sizes for better readability and accessibility.
\end{itemize}
 So, the final experimental sample was determined by selecting colors and their count, font size, and logo position. Finally, after the procedure, the following representative Kansei web design was created (see Fig. \ref{finfig}).
 
Collaboration with the company's designers resulted in creating a beautiful new website design. Using the PCA findings helps align the designs with the most relevant Kansei principles and enables a targeted collaboration with designers. The design process can be finely tuned by iterating designs with these factors in mind, possibly complemented by follow-up user surveys to test new concepts.

In this section, we demonstrated how to integrate user emotions into the design of web products using KE. While this study focused on the Reis.kz platform as a specific case, the presented approach can be applied in similar industries, like e-commerce websites, social media, and educational platforms, and many other online environments where emotional responses are important.
\section{Discussion}
Let us see how the current study’s findings compare to previous studies. 

Recent study\cite{webpage1} describes a website homepage that employed KE. Their main idea is to optimize a method to design products that closely match customers' emotional preferences. Similar studies\cite{discussion5}, \cite{mobilek} show that embedding KE into mobile application design can result in a user-friendly and intuitive interface. Another work\cite{Liu2023} has been conducted in the past. However, it focused on gathering Kansei's words by analyzing a large number of product reviews available online. 

We emphasize that we are investigating the integration of Kansei Engineering into the transportation company's website. We invited participants from different fields at different stages of the research to identify and select the most important design elements. Also, we can compare by main website features (e.g., font, color, image ratio) that were identified as primary in other studies\cite{Kandambi2022}, \cite{discussion4}, \cite{discussion2} as well. Our findings confirm previous research studies\cite{discussion1}, \cite{discussion3}, which discovered website interfaces focusing on user preferences. 

This research's findings contribute to the growing body of knowledge on KE and its application in web interface design.\\

\section{Future Works and Limitations}

This research has several limitations. One primary limitation was the reliance on website images for Kansei criteria evaluation. We mainly used website images to see how people feel about it. This method might not show us everything about how users really experience the website when they use it. If participants could interact directly with the website instead of just looking at pictures,   additional insights could be revealed that static images cannot provide, potentially leading to a more accurate evaluation of Kansei attributes.

In future research, we plan to involve more respondents to get a broader view of different users' feelings about certain designs. 
Additionally, incorporating more product samples could allow for a more detailed exploration of product feature-emotion associations. Further research could also explore the effects of direct user interaction with the website compared to evaluating static images.
In addition, alongside PCA, other techniques, such as Factor Analysis and ML algorithms like Random Forests or Support Vector Machines, could also be employed to classify emotional responses more accurately.

\section{Conclusion}


In this paper, we utilized the Kansei approach to analyze the emotional impact on the logistics website, a platform for connecting cargo services and clients. Our study involved surveys and experiments to link specific emotional responses to website features, which were then integrated into the design to enhance user experience. By applying KE to the transportation sector, which is traditionally not associated with aesthetic focus, we explored its potential to improve web design beyond conventional applications like fashion or beauty industries. The proposed approach produced a web design integrated into a real web application.

The research findings have practical implications for designers, developers, and web interface design experts. By addressing users' emotional responses, designers can create interfaces that fulfill functional requirements and create positive and meaningful experiences. 



\section*{Acknowledgements}
 The authors extend their appreciation to the team at Reis.kz for their support and cooperation throughout the study.

\bibliography{sn-bibliography}

\end{document}